\documentclass[aps,prl,preprint,showpacs,preprintnumbers,amsmath,amssymb,groupedaddress]{revtex4-1}
\usepackage{graphics}
\usepackage{graphicx}
\usepackage{epsfig}
\usepackage{amsmath}
\usepackage{bm}
\begin{document}
\title{Surface Roughness Dominated Pinning Mechanism of Magnetic Vortices in Soft Ferromagnetic Films}
\author{T. Y. Chen}
\affiliation{School of Physics and Astronomy, University of Minnesota, 116 Church St. SE, Minneapolis, MN 55455}
\author{M. J. Erickson}
\affiliation{School of Physics and Astronomy, University of Minnesota, 116 Church St. SE, Minneapolis, MN 55455}
\author{C. Leighton}
\affiliation{Department of Chemical Engineering and Materials Science, University of Minnesota, 421 Washington Ave. SE, Minneapolis, MN 55455}
\author{P. A. Crowell}
\affiliation{School of Physics and Astronomy, University of Minnesota, 116 Church St. SE, Minneapolis, MN 55455}

\begin{abstract}
Although pinning of domain walls in ferromagnets is ubiquitous, the absence of an appropriate characterization tool has limited the ability to correlate the physical and magnetic microstructures of ferromagnetic films with specific pinning mechanisms. Here, we show that the pinning of a magnetic vortex, the simplest possible domain structure in soft ferromagnets, is strongly correlated with surface roughness, and we make a quantitative comparison of the pinning energy and spatial range in films of various thickness. The results demonstrate that thickness fluctuations on the lateral length scale of the vortex core diameter, i.e. an effective roughness at a specific length scale, provides the dominant pinning mechanism. We argue that this mechanism will be important in virtually any soft ferromagnetic film.
\end{abstract}

\maketitle

The pinning of domain walls in ferromagnets is attributed to the interactions between the domain structure and local fluctuations of magnetic properties due to defects. Possible sources of defects in polycrystalline ferromagnets include point defects (e.g. impurities, vacancies, and nonmagnetic inclusions), line defects (e.g. dislocations), surface imperfections (e.g. roughness), and random anisotropies. Because multiple types of defects coexist in a given material, it is problematic to identify exactly which ones dominate the pinning process. It has been practically impossible to identify individual pinning sites, and therefore studies of domain wall pinning have focused on collective effects \cite{brice_easy_1966, dijkstra_effect_1950, herzer_grain_1990, li_effect_1998}. This approach is not adequate for applications of domain wall-based devices in which pinning must be precisely engineered. An alternative approach is to study pinning in simple, albeit non-uniform magnetic structures, where individual pinning sites can be readily identified. Recent studies of single magnetic vortices in ferromagnetic disks provide excellent examples, showing discontinuous vortex motion as a function of the applied magnetic field \cite{uhlig_shifting_2005}, defect-induced enhancement of the gyrotropic frequency \cite{compton_dynamics_2006, compton_magnetic_2010, kim_current-induced_2010}, and non-linear vortex dynamics due to anharmonic pinning potentials \cite{chen_2010}. It is thus clear from these studies that pinning of a single vortex can in fact be probed via vortex dynamics, although the precise pinning mechanism has remained largely mysterious.

In this Letter, we report on the dominant pinning mechanism for a single magnetic vortex in soft ferromagnetic permalloy (Ni$_{80}$Fe$_{20}$) films, quantifying pinning energies and spatial ranges as a function of film thickness. We show that the measured pinning range, approximately 20 nm, is nearly identical to the vortex core diameter, demonstrating that the pinning defects interact only with the core of the vortex. We further show, using the thickness dependence of the pinning energy, that the dominant pinning defects are located on the surfaces. We demonstrate quantitatively that the pinning is correlated not with the root mean square (RMS) of the surface roughness but rather with the roughness on the lateral length scale of the core diameter, i.e. an effective roughness. We argue that this vortex-pinning mechanism will be important in virtually any soft ferromagnetic film. Our findings are thus directly relevant to magnetic devices containing vortices, such as writer poles in hard disk drives \cite{patwari_simulation_2010}, magnetic nanowires with vortex domain walls \cite{parkin_magnetic_2008}, and vortex-type spin-torque oscillators \cite{pribiag_magnetic_2007}.

Stable vortices are obtained in micron size ferromagnetic disks in which the magnetization curls in the plane of the disk, with the exception of the disk center, where the magnetization orients out of the plane within a core region on the order of 10 nm in diameter. In our case the magnetic disks were patterned from polycrystalline permalloy (Ni$_{80}$Fe$_{20}$) films, which were grown on Si substrates with a SiN buffer layer by dc magnetron sputtering at 100 W (0.1 nm/s) in 2.5 mTorr Ar, at ambient temperature. The film thicknesses studied were 20, 35, 50, 65, 80, 100, and 130 nm. On each permalloy film, a 70 nm thick Ti layer was deposited as a hard mask, which was patterned into 1-$\mu$m-diameter disks using electron beam lithography. The disk patterns were then transferred to the Ni$_{80}$Fe$_{20}$ layer by Ar ion milling.
\begin{figure*}
\centerline{\epsfbox{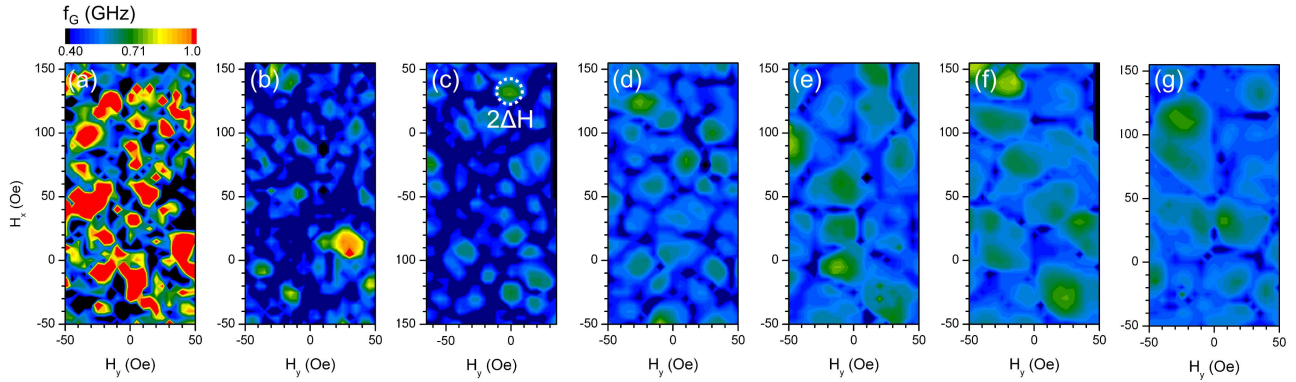}}
\caption{(Color online) Contour maps of the gyrotropic frequency $f_G$ as a function of the in-plane static field for disk thicknesses of (a) 20 nm, (b) 35 nm, (c) 50 nm, (d) 65 nm, (e) 80 nm, (f) 100 nm, and (g) 130 nm. High frequency areas in the contour maps correspond to pinning sites, as indicated by the dashed circle in (c) for example. $\Delta H$ is the radius of a pinning site as defined in the text.}
\label{fig:fig1}
\end{figure*}

We used time-resolved Kerr microscopy (TRKM) to measure the vortex gyrotropic mode of individual magnetic disks, i.e. the lowest frequency excitation of the vortex \cite{park_imaging_2003}. We first mapped the spatial distribution of pinning defects by measuring the gyrotropic frequency $f_G$ as a function of the orthogonal in-plane d.c. magnetic fields \cite{compton_dynamics_2006, compton_magnetic_2010}. The in-plane fields were varied in increments of 5 Oe over a range of 100 Oe $\times$ 200 Oe, which displaces the vortex core over a 110 $\times$ 220 nm$^2$ spatial region around the center of the disk. For each set of static magnetic field values, the gyrotropic mode was excited by a magnetic field pulse with a temporal width less than 120 ps and an amplitude of 5 Oe oriented in the plane of the disk, and the resulting gyrotropic frequency $f_G$ was measured. Contour maps of $f_G$ as a function of the static fields are shown in Fig.~\ref{fig:fig1}. $f_G$ is represented by a color scale, and pinning sites appear as localized regions of high $f_G$.

\begin{figure}
{\epsfbox{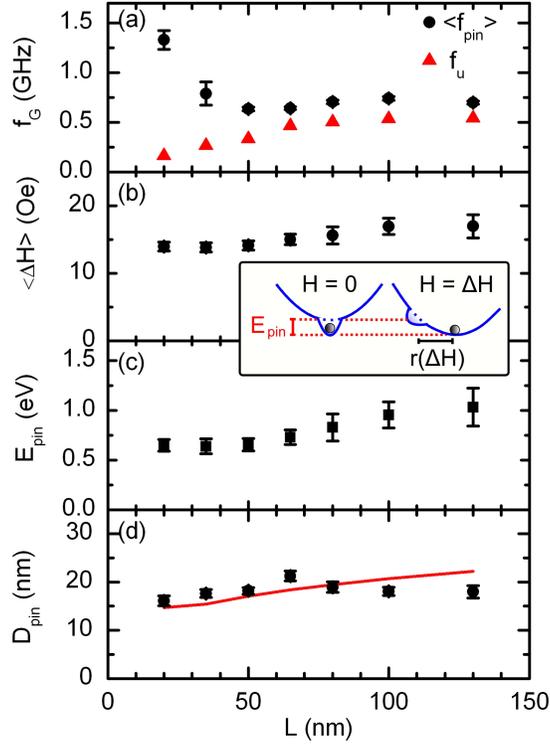}}
\caption{(Color online) Thickness dependence of (a) the pinned ($f_{pin}$) and the unpinned ($f_u$) gyrotropic frequency, (b) the averaged depinning field $\Delta H$, (c) the pinning energy $E_{pin}$, and (d) the pinning range $D_{pin}$. The solid line in (d) is the size of core obtained from the micromagnetic simulation.  The inset shows a schematic of how the total potential energy changes in an applied field.}
\label{fig:fig2}
\end{figure}

We characterized each pinning site in Fig.~\ref{fig:fig1} via two quantities, the pinned frequency $f_{pin}$, which is the highest frequency within each point-like area, and the depinning field $\Delta H$, where $2\Delta H$ is the full width at half maximum (FWHM) of the $f_G$ peak, averaged from the two orthogonal field directions. The averaged pinning-site characteristics ($\langle f_{pin}\rangle$ and $\langle \Delta H \rangle$) for each sample are shown in Figs.~\ref{fig:fig2}(a) and (b) respectively. It is clear that $\langle f_{pin}\rangle$ is significantly higher in thinner disks, while $\langle\Delta H\rangle$ is relatively insensitive to thickness $L$.

Also shown in Fig.~\ref{fig:fig2}(a), is the ``unpinned" gyrotropic frequency $f_u$, which is obtained by measuring $f_G$ at higher amplitude to remove the influence of pinning \cite{chen_2010}.   $f_u$ increases with $L$ as expected from analytical models and micromagnetic simulations \cite{guslienko_eigenfrequencies_2002, park_imaging_2003, park_interactions_2005, v._novosad_magnetic_2005}. It is surprising however, that the enhancement of $f_G$ due to pinning ($\langle f_{pin}\rangle-f_u$) varies approximately as $1/ L$. As will be discussed below, the enhancement of $f_G$ is associated with the lateral range of pinning, while the $1/L$ trend reflects the existence of a single length scale characterizing the pinning interaction.

Within a simple model, the measured $\Delta H$ and $f_{pin}$ allow us to determine the physical properties of each pinning site, including the pinning energy $E_{pin}$ and the pinning range $D_{pin}$. This can be done within the ``two-vortices" model \cite{guslienko_eigenfrequencies_2002} developed to describe vortex dynamics in a disk, considering the local pinning potential created by a defect. The core is assumed rigid, and the geometric confinement $W$ is approximated by a parabolic potential, $W(r)=k_{u}r^2/2$, where $r$ is the distance between the core and the disk center, $k_{u} = M_s^2 L \xi^2 \pi /\chi_0$ is the unpinned stiffness, $M_s$ is the saturation magnetization, $\chi_0 = R/10L$ is the vortex susceptibility, and $\xi$ = 2/3 is a model-dependent constant. The applied in-plane magnetic field $H$ changes the potential energy by $H\mu(r)$, where $\mu(r)=\xi M_s \pi r RL$, and $R$ is the radius of the disk. Thus, in an applied field the core moves to a new equilibrium position $r(H)=\chi_0 RH/(M_s\xi)$ for an unpinned vortex. In contrast, for a pinned vortex the core is trapped by a local pinning potential $W_{pin}(r)$, unless a sufficiently large magnetic field $\sim$ $\Delta H$ is applied to overcome the energy barrier. This is illustrated in the inset of Fig.~\ref{fig:fig2}.  For a pinning site in the middle of the disk, we can estimate the pinning energy $E_{pin}$, i.e. the energy barrier, from $\Delta H$ using
\begin{equation}
E_{pin}=\frac{1}{2} k_u r(\Delta H)^2.
\label{eq:eq1}
\end{equation}
This can be generalized to a pinning site at any location in the disk by shifting the origin. As shown in Fig.~\ref{fig:fig2} (c), $\langle E_{pin}\rangle$ is constant at small thickness and then drifts slightly upward with $L$. If the pinning effect were dominated by point defects in the bulk of the film, we would expect the pinning energy to scale with the number of defects intercepted by the core as it traverses the film.  The pinning energy would then be proportional to $L$, which is not observed. Rather, the thickness dependence suggests that the observed vortex pinning is dominated by surface defects, motivating a detailed analysis of surface roughness as provided below.

To determine $D_{pin}$, we consider the simplest form of a pinning potential, $W_{pin}(r)=\Delta k r^2/2$, where $\Delta k\equiv k_{pin}-k_u =2\pi (f_{pin}-f_u)G$, and $G=2 \pi L M_s/\gamma$ is the gyroconstant \cite{guslienko_eigenfrequencies_2002}. We define $D_{pin}$ as the diameter of the pinning potential, so that $W_{pin} (D_{pin}/2)=E_{pin}$, and
\begin{equation}
D_{pin}=2\sqrt{\frac{E_{pin}}{\pi (f_{pin}-f_u)G}}.
\label{eq:eq2}
\end{equation}
As shown in Fig.~\ref{fig:fig2}(d), $D_{pin}$ approximately matches the diameter of the vortex core, which is indicated by the solid curve. The core diameter is obtained from a micromagnetic simulation using typical material constants for permalloy, including $M_s=800$~emu/cm$^3$ and the exchange constant $A=1.05 \times 10^{-6}$~ergs/cm.  Because the core diameter at the disk surface is smaller then in the equatorial plane of the disk \cite{fischer_x-ray_2011}, in the simulation we divide the disk into 5 layers, and we define the core diameter by the radius of the maximum gradient of the \emph{z}-component magnetization at the surface of the disk. It is likely that local defects pin the vortex core region more effectively than the other regions of the vortex due to the large energy density within the core, where the magnetization gradient is large. Imperfections on length scales similar to the core diameter therefore cause the strongest pinning effects.

It should be noted that the above explanation for the observed pinning range is consistent with the consensus based on models of domain wall pinning \cite{alex_magnetic_1998}. Specifically, pinning of domain wall motion is known to be most effective at defects with dimensions comparable to the wall width. However, to our knowledge this limiting-defect-size effect has been shown in experiments only through collective effects \cite{dijkstra_effect_1950,ohandley_magnetization_1985, herzer_grain_1990}, in which the highest coercivities were observed when the grain or inclusion size in the film matched the estimated domain wall width. Here, because we directly identify the spatial range of the interaction between a single vortex and an individual pinning site, which is set by the core diameter (Fig.~\ref{fig:fig2}(d)), our findings represent strong evidence for the predicted limiting-defect-size effect.

\begin{figure}
\centerline{\epsfbox{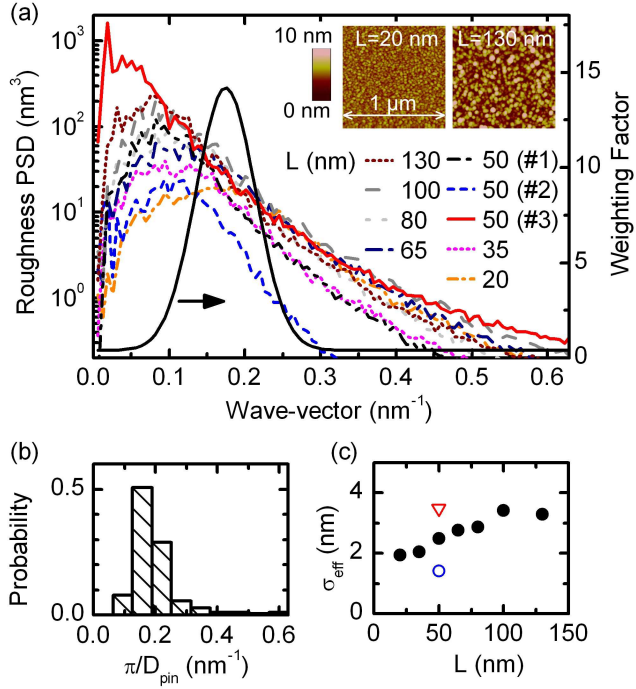}}
\caption{(Color online) (a) Roughness power spectral density (PSD) as a function of wave-vector (lines), obtained from the Fourier transforms of atomic force micrographs (AFM).  (b) Probability distribution of the pinning range as a function of $\pi/D_{pin}$. Data are obtained from the pinning sites in Fig.1. This distribution is used to determine the weighting function shown by the solid Gaussian curve in (a).  (c) Effective roughness $\sigma_{eff}$ versus the disk thickness $L$.  The two points shown in open symbols are additional 50 nm thick films prepared with different roughness characteristics.}
\label{fig:fig3}
\end{figure}

We now turn to analysis of the surface roughness, which was characterized using tapping-mode atomic force microscopy (AFM). Representative AFM images of the 20-nm-thick and 130-nm-thick samples are shown in the inset of Fig.~\ref{fig:fig3}(a). The grain size and roughness are found to increase with thickness, as is typical. This can be seen in Fig.~\ref{fig:fig3}(a), which shows the roughness power spectral density (PSD). By integrating the spectra, we can obtain the RMS value of the roughness $\sigma_{RMS}$.  To determine the contribution to the roughness from length scales on the order of the core diameter, we integrate the roughness PSD with a weighting function, shown as the Gaussian solid curve in Fig.~\ref{fig:fig3}(a).  This function is determined experimentally from the probability distribution of $\pi/D_{pin}$, which is shown in Fig.~\ref{fig:fig3}(b).  We use $\pi/D_{pin}$ as the relevant length scale because the vortex core will be accommodated most easily when fluctuations in the pinning potential are twice its diameter.  Using this procedure, we obtain an effective roughness $\sigma_{eff}$ for each sample, as shown in Fig.~\ref{fig:fig3}(c). $\sigma_{eff}$ increases from approximately 2 to 4 nm as $L$ increases from 20 to 130 nm.

\begin{figure}
\centerline{\epsfbox{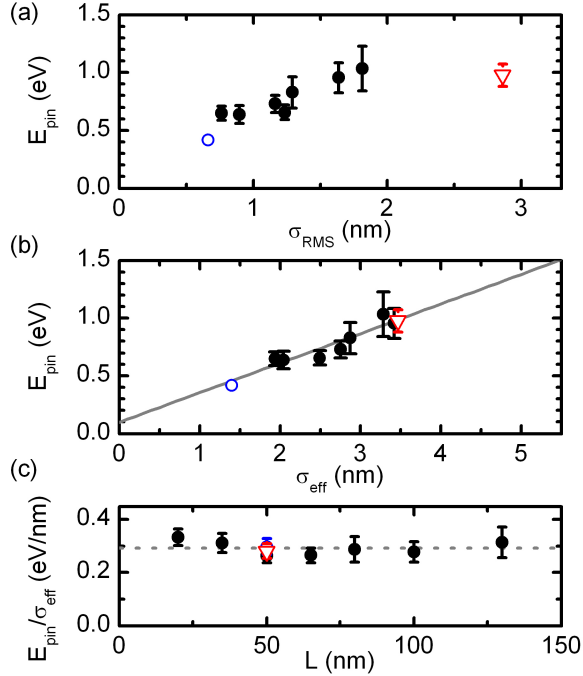}}
\caption{(Color online) Pinning energy versus (a) the RMS roughness, and (b) the effective roughness (on the length scale of the core diameter). The solid line is a linear fit. (c) Pinning energy normalized by the effective roughness versus disk thickness. Data obtained on the 50 nm thick sample \#2 are indicated by open circles. Data obtained on the 50 nm thick sample \#3 are indicated by open triangles. The dashed line indicates the average value of $E_{pin}/\sigma_{eff}$.}
\label{fig:fig4}
\end{figure}

The relationship between pinning and roughness is shown in Figs.~\ref{fig:fig4}(a) and (b), which show $E_{pin}$ as a function of $\sigma_{RMS}$ and $\sigma_{eff}$, respectively. Besides the samples discussed thus far (filled circles), two additional 50 nm thick samples are shown for comparison, indicated by open circles for 50 nm sample \#2 and open triangles for 50 nm sample \#3. Sample \#2 has smaller roughness compared to the original 50 nm sample (\#1), and was prepared in a different deposition run. Sample \#3 has larger grain size ($\sim$100 nm) than samples \#1 and \#2 ($\sim$30 nm), as it was deposited at an elevated substrate temperature of 250 C \cite{compton_magnetic_2010}. As can be seen in Fig.~\ref{fig:fig4}(a), there is no clear linear correlation between $\sigma_{RMS}$ and $E_{pin}$. However, as shown in Fig.~\ref{fig:fig4}(b), it is clear that $E_{pin}$ scales linearly with $\sigma_{eff}$, and the linear fit, shown as a solid line, intercepts with the y-axis approximately at zero. This result indicates that the effective roughness $\sigma_{eff}$ is the dominant vortex pinning mechanism for all of these permalloy films. The correlation between $E_{pin}$ and $\sigma_{eff}$ also explains the thickness dependence of $E_{pin}$, shown in Fig.~\ref{fig:fig2}(c). This is due to the fact that $\sigma_{eff}$ is larger for thicker disks (Fig.~\ref{fig:fig3}(c)), as evidenced by the constant value of $E_{pin}/\sigma_{eff}$ versus thickness, as shown in Fig.~\ref{fig:fig4}(c).

We can estimate how small the effective roughness would need to be in order to avoid the roughness-induced pinning mechanism. We consider that a pinning site is unimportant only if depinning of a vortex from that site can be thermally activated at room temperature on a typical laboratory time scale, i.e. 1 sec. With an attempt frequency of 0.5 GHz, which is set by the gyrotropic frequency, the critical pinning energy turns out to be 0.5 eV. Therefore, from the data in Fig.~\ref{fig:fig4}(b), we determine that the effective roughness would need to be smaller than 1.5 nm to avoid vortex pinning.  To put this effective roughness value in perspective, we consider its implication for films with various growth modes. In the Volmer-Weber or island growth mode (often relevant to sputtered polycrystalline metal films), roughness is significant and the lateral correlation length is directly linked to the grain size. Given that this length scale in most practical situations is of the same order of magnitude as the physically relevant pinning length scale (i.e. 20~nm, the vortex core diameter), the constraint  $\sigma_{eff} \ll 1.5$~nm is a very stringent one, requiring grain sizes very different from the core diameter. In Frank-van der Merwe (i.e. layer-by-layer) or step-flow growth modes (potentially relevant to MBE-grown epitaxial metal films) the lateral correlation length of the roughness is set by the mean terrace width, and thus the vicinality of the substrate surface. Satisfying  $\sigma_{eff} \ll 1.5$~nm may be possible, but even in this case it would require specific tailoring of the vicinality and terrace width to avoid the scale of the core diameter. Thus, we expect that the surface roughness pinning mechanism plays an important role for vortex pinning in virtually all soft ferromagnetic films. Similar length scale arguments apply to edge roughness in patterned thin film devices, as the length scale associated with the patterning technique, such as electron beam lithography, is again likely on a similar scale to the vortex core.

In conclusion, we have shown that the dynamics of a single vortex allow us to quantify both the energies and length scales associated with individual pinning sites. The dominant pinning mechanism is the interaction between the vortex core and surface roughness on a lateral length scale set by the core diameter. We suggest that this  mechanism determines the minimum pinning energy for vortex motion in most soft ferromagnetic thin-film devices, such as vortex domain walls in nanowires.

This work was supported primarily by the MRSEC Program of the National Science Foundation under Award Number DMR-0819885. Additional support for the Nanofabrication Center was provided by the NSF NNIN network.

\end{document}